\documentclass[10pt]{article}
\usepackage{ae}
\usepackage{aecompl}       
\usepackage[latin1]{inputenc}
\usepackage{geometry}
\usepackage{setspace}
\usepackage{graphics} 
\usepackage{verbatim}
\usepackage{amsmath}
\usepackage{url}

\usepackage[authoryear, round, sort]{natbib}

\title{Molecular Signatures from Gene Expression Data} 
\author{Ramón Díaz-Uriarte \\
Bioinformatics Unit\\
Spanish National Cancer Center (CNIO)\\
Melchor Fernández Almagro 3 \\
Madrid, 28029\\
Spain. \\
\texttt{rdiaz@cnio.es}\\
\url{http://ligarto.org/rdiaz}\\
}

\date{
\vspace*{60pt}
Version 3, September 2004 \\
\vspace*{40pt}
{\bf Running Head:} Gene expression signatures.} 

\begin{document}

\maketitle

\newpage

\begin{abstract}

  {\bf Motivation} ``Molecular signatures'' or ``gene-expression signatures''
  are used to predict patients' characteristics using data from coexpressed
  genes. Signatures can enhance understanding about biological mechanisms and
  have diagnostic use. However, available methods to search for signatures fail
  to address key requirements of signatures, especially the discovery of sets
  of tightly coexpressed genes.
  
  {\bf Results} After suggesting an operational definition of signature, we
  develop a method that fulfills these requirements, returning sets of tightly
  coexpressed genes with good predictive performance. This method can also
  identify when the data are inconsistent with the hypothesis of a few, stable,
  easily interpretable sets of coexpressed genes.  Identification of molecular
  signatures in some widely used data sets is questionable under this simple
  model, which emphasizes the needed for further work on the operationalization
  of the biological model and the assessment of the stability of putative
  signatures.

  {\bf Availability} The code (R with C++) is available from 
  \url{http://www.ligarto.org/rdiaz/Software/Software.html} under
  the GNU GPL.

{\bf Contact} \texttt{rdiaz@cnio.es}

{\bf Supplementary information}
 \url{http://ligarto.org/rdiaz/Papers/signatures-supl.mat.pdf}\\

\end{abstract}

\section{Introduction}

``Molecular signatures'' or ``gene-expression signatures'' are a key feature in
many studies that use microarray data in cancer research 
\citep[e.g.,][]{alizadeh,golub,rosenwald,shipp,pomeroy}.  
\citet[p.~375]{shaffer}
refer to signatures as ``(...) genes that are \textbf{coordinately expressed}
in samples related by some identifiable criterion such as cell type,
differentiation state, or signaling response'' (emphasis is ours).  Molecular
signatures are often used to model patients' clinically relevant information
(e.g., prognosis, survival time, etc) as a function of the gene expression
data, but instead of using individual genes as predictors, the predictors are
the signature components or ``metagenes''.

If we are successful searching for a signature, then we will be able to model,
for instance, the probability of developing a metastasis as a function of a few
signature components or metagenes where each signature component is made of
genes that show strong coexpression. Thus, molecular or gene expression
signatures can be important both for diagnostic purposes and
for providing information about the biological mechanisms underlying certain
conditions by highlighting genes that both coexpress and are related to that
condition.

\begin{figure}[th!]
\begin{center}
{\resizebox{\columnwidth}{!}{%
\includegraphics{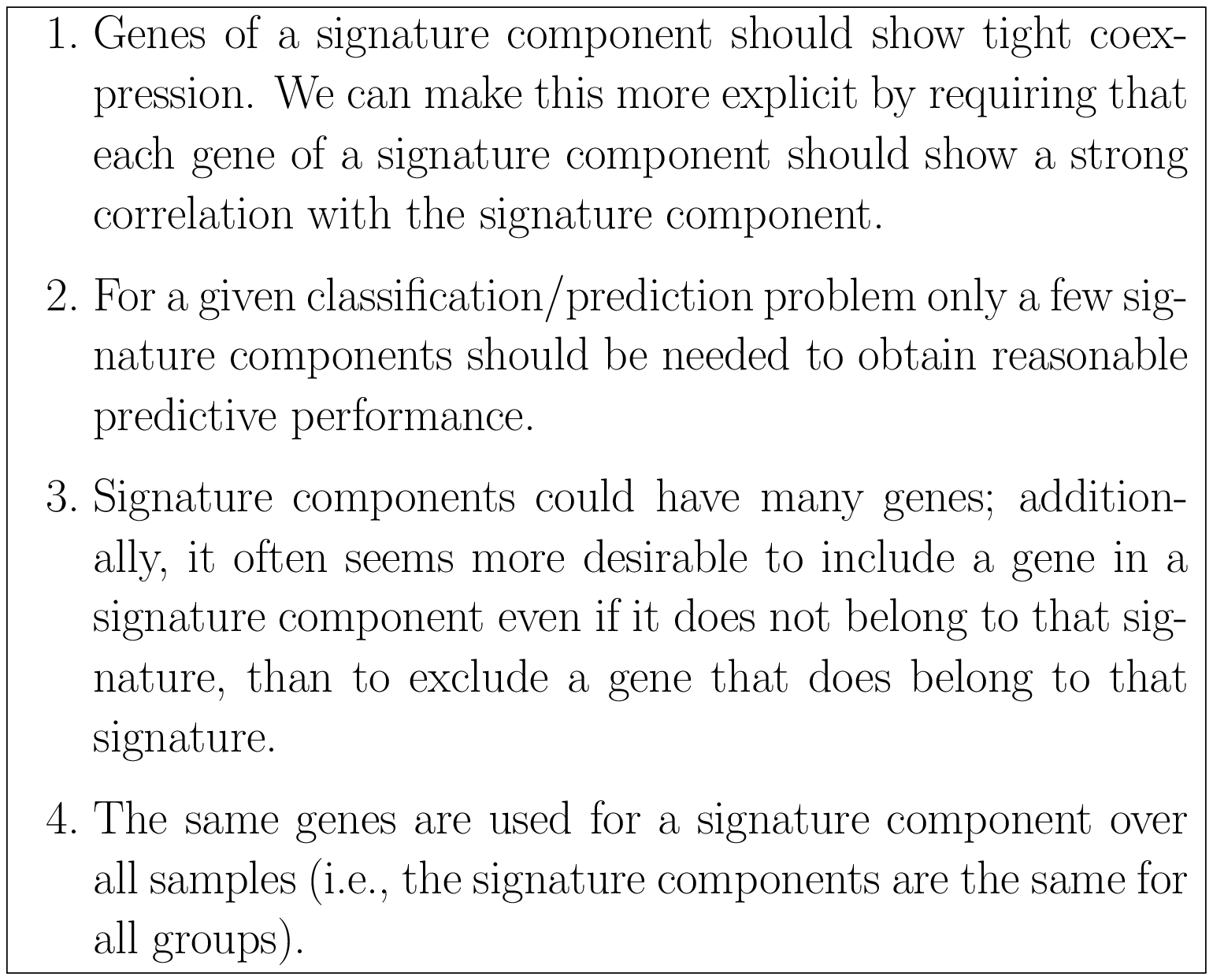}}}
\caption{\label{box-signature} Requirements of signatures and
  signature components (see text for details).\hspace{200pt}}
\end{center}
\end{figure}

In spite of the widespread use of the term ``molecular signature'', no explicit
definition is available. Following the conventions of the literature
\citep[e.g.,][]{alizadeh, golub, rosenwald, shipp, pomeroy, West-pnas}, and
building upon the definition above \citep[][p.~375]{shaffer}, we will consider
a signature to be composed of one or more \textbf{signature components} or
\textbf{metagenes}, where each signature component is a weighted combination of
one or more coexpressed genes, and such that statistical models that use
signatures both have good predictive performance and are easy to interpret
biologically.  Interpretation is made easier because the prediction is based on
signature components that are weighted averages of \textbf{subsets of tightly
  coexpressed genes}, which can help when attempting to relate specific
biological features to, for example, particular alterations on a metabolic
pathway.  Based upon the above references, we can try to formalize these goals
by requiring that signatures and signature components satisfy the conditions
shown in Figure \ref{box-signature}.

\vspace{20pt}
\centerline{[Figure 1 about here]}
\vspace{20pt}

The conditions in Figure \ref{box-signature} reflect a very specific biological
model. Our objective is to develop a statistical method appropriate for this
biological model. By using a method that tries to fulfill those conditions we
can also provide evidence that, for any particular case, the underlying
biological assumptions behind this attempt are inconsistent with the data or,
in other words, the the assumptions embodied in Figure \ref{box-signature} are
inappropriate. As will be discussed later, our method is an attempt to map a
particular biological model into a statistical method, but other statistical
approaches would be more appropriate if more complex biological models are
regarded as appropriate.

\subsection{Limitations of alternative methods}

A variety of approaches have been used to identify molecular signatures. A
review is provided in the supplementary material.  Briefly, most methods
return either a single signature component \citep[e.g.,][]{golub, hedenf, vveer}
which is a weighted average of a set of genes, or several signature components
\citep[e.g.,][]{West-pnas, ramas-svm, harvest, huang-nature, mave-dna} which are
often obtained using dimension reduction techniques (e.g., principal component
analysis [PCA], partial least squares [PLS], sufficient dimension reduction)
either on the complete set of genes or on a preselected subset.

The most common problems of available methods are:
\begin{itemize}
\item Genes within signature components do not necessarily show tight
  coexpression: no method makes tight coexpression a requirement to be
  fulfilled. 

\item The interpretation of components is very difficult for most
methods that use PCA or PLS, since all the genes to which PCA or PLS
is applied have loadings on each component.

\item The search for components in many PCA or clustering of genes
methods is carried out without incorporating information from the
dependent variable.

\item Most methods are designed for a specific type of task (e.g.,
classification or survival, but not both) and would be difficult to
extend to other types of dependent variables.
\end{itemize}

Our objective in this paper is to propose a method that overcomes these
problems. Specifically, our method returns signature components of tight
coexpression (and thus, signature components that should ease interpretation)
with good predictive performance.  Although in this paper we focus on class
prediction, our method can be used with different types of dependent data
(continuous, categorical, survival), and thus sets up a general framework for
finding gene expression signatures regardless of the type of dependent
variable.

Based on the proposed operational definition of signatures (see figure
\ref{box-signature}), we first discuss the key elements of our proposed method.
Next, we evaluate its predictive performance and finally with discuss it relation with
other methods and problems of biological interpretability.  Further details
about the algorithm, evaluation of recovery of signatures, and a longer review
of alternative approaches are provided in the supplementary material.

\section{Methods}

\subsection{Key elements of the proposed method}
Our objective is to directly fulfill the
conditions in figure \ref{box-signature}.  We start our search with a seed gene
that will be the skeleton of a signature component; this first signature
component is found so that genes within the component show tight coexpression
and the prediction error is acceptable. We repeat this process (find seed gene
for a component and then obtain the whole component) greedily, until no further
components are needed.  The main steps of the algorithm are shown in Figure
\ref{fig-algorithm}. In this section we explain how the conditions in figure
\ref{box-signature} can be fulfilled and provide a geometrical interpretation
of the algorithm.

\vspace{20pt}
\centerline{[Figure 2 about here]}
\vspace{20pt}

\subsubsection{\label{fulfill}Fulfilling signature requirements}

A common and simple way of characterizing a signature component is to use
linear combinations (weighted averages) of the genes that belong to that
signature component \citep[e.g.,][]{West-pnas, shipp, ramas-03, rosenwald}.
Although we could characterize a signature component using several different
linear combinations of the genes of that component, most methods \citep[but
see][]{liu-pca} characterize a signature component or metagene using only one
linear combination.  A single metagene per signature component simplifies
interpretation, and is implicit in the requirement that each gene of a
signature components should show a strong correlation with the signature
component.

Thus, to fulfill requirement one in figure \ref{box-signature}, we can use
Principal Component Analysis (PCA ---which is closely related to Singular Value
Decomposition [SVD]). PCA yields ``the best'' representation (or ``least
distorting'' representation, in the least squares sense) of the original data
\citep[e.g.,][]{jolliffe, krza-book, morrison} in a subspace of reduced
dimensions. The first PC is the best 1-dimensional representation of the
original genes of this signature component. If the genes of the signature
component are tightly coexpressed, then each of these genes should show a high
correlation with the signature component, as we required above (this will also
mean that the percentage of variance in the original gene expression data
explained by this first PC will be high ---see supplementary material). After
running the procedure, each signature component will be made of tightly
coexpressed genes (we require that all the genes in a component show
a correlation larger than a pre-specified threshold of $r_{min}$).

In contrast to some previous methods which use PCA or PLS over the complete set
of genes, there is no need for our method to return components which are
uncorrelated or orthogonal: there is no biological argument that requires that
signature components be orthogonal, uncorrelated, or independent (see
discussion).  For ease of interpretation we will additionally require that no
gene belongs to more than one signature component.  (In other words, each gene
in the original data matrix belongs to either one, and only one, signature
component or to none.)

The second and third requirements of figure \ref{box-signature} can be
incorporated by adding new signature components only if they result in a
relevant reduction of prediction error, and retaining genes in a signature
unless they produce large increases in prediction error. In other words, we
will penalize adding signature components, but will try to obtain large
signature components. The reason is that the search for molecular signatures is
often pursued to provide biological insights into coexpressed genes related to
conditions and, therefore, minimization of prediction error is not the only
goal. Thus, if there are potential trade-offs between prediction error and
``biologically interpretable signatures'', the researcher should have the
option of modifying the terms of this trade-off flexibly.

\subsubsection{\label{geometry}Searching for signature components and a geometrical interpretation}

Our objective is, thus, to maximize predictive performance using signature
components that satisfy that the correlation of each gene in a signature
component with the signature component is larger than a given threshold.
However, the discussion so far does not indicate how to find the signature
components and, given the dimensionality of the problem, an exhaustive search
for the optimal solution is not possible.  Since we require that each component
be highly correlated with the genes of that component, we can start the search
with genes that have good predictive abilities on their own. Once we find an
initial \textbf{``seed gene''}, we build an initial candidate signature
component by including all ``promising genes'' (e.g., all those with a minimum
correlation with the seed gene), and later reduce the signature component
eliminating genes until the conditions of minimum correlation with 1st PC (all
genes have a correlation with the 1st PC $> r_{min}$) and predictive
performance are met. (If this elimination eliminates all genes except the seed
gene, then, of course, the two requirements are met).

The method proposed here is a heuristic search that has an intuitive
geometrical interpretation. We require that each component be highly correlated
($> r_{min}$) with the genes in the component, which is equivalent to saying
that the vector of the component must have a similar direction as the vectors
of each gene in variable space (the space where subjects are the axes).
Therefore, no matter which genes belong to a signature component, the component
will have a similar direction as any of its genes. Then, it seems reasonable to
start the search with the direction that has the best predictive ability, the
seed gene; this seed gene is the single direction in space that most
contributes to separation of the groups in a classification problem; analogous
for regression or survival analysis.  When we form the complete signature
component, all other genes of the signature component have directions that are
similar to that of the seed gene. Together, all the genes of a signature
component move the direction slightly, but this shift is possibly towards
directions that contribute more to separation of groups (or that at least do
not degrade the separation too much) and never moves us far away from the
original seed gene. This process is repeated until the addition of new
signature components does not achieve any relevant decrease in prediction rate,
or until a maximum pre-specified number of signature components is reached.
The algorithm is shown in figure \ref{fig-algorithm}. Further details are given
in the supplementary material.

\subsubsection{Choice of underlying classifier}

In this paper we will be dealing with a classification problem. Each signature
component is used as a predictor variable for a classifier.  Of the available
classification methods, we have used DLDA (diagonal linear discriminant
analysis), a version of linear discriminant analysis which assumes the same
diagonal variance-covariance matrix for all the classes \citep{dudoit-dlda}, and
NN (k-nearest neighbor, with k = 1), a simple non-parametric rule that assigns
a test sample to the class of the closest training sample (where closeness is
measured using Euclidean distance in the space whose dimensions are the
signature components). KNN and DLDA have been repeatedly shown to perform as
well as, or better than, many competing methods with microarray data
\citep{dudoit-dlda, romualdi-03}. In addition, DLDA and KNN are simple to
implement and interpret. \cite{dudoit-dlda} used an adaptive procedure to
estimate the optimal number of neighbors to use with KNN; that can be time
consuming, and we have fixed K = 1, since this is often a successful rule
\citep{ripley-96, htf-01}.  As discussed in the supplementary material, other
classifiers can be used.

\section{Comparing predictive performance with established methods}

Here we compare the predictive performance of our method with that of three
well established methods, support vector machines, KNN, and DLDA, using several
``real data'' sets.  The supplementary material reports simulation studies that
show that the suggested method can indeed recover signatures when they are
present in the data.

\begin{figure}[h!]
\begin{center}
{\resizebox{\columnwidth}{!}{%
\includegraphics{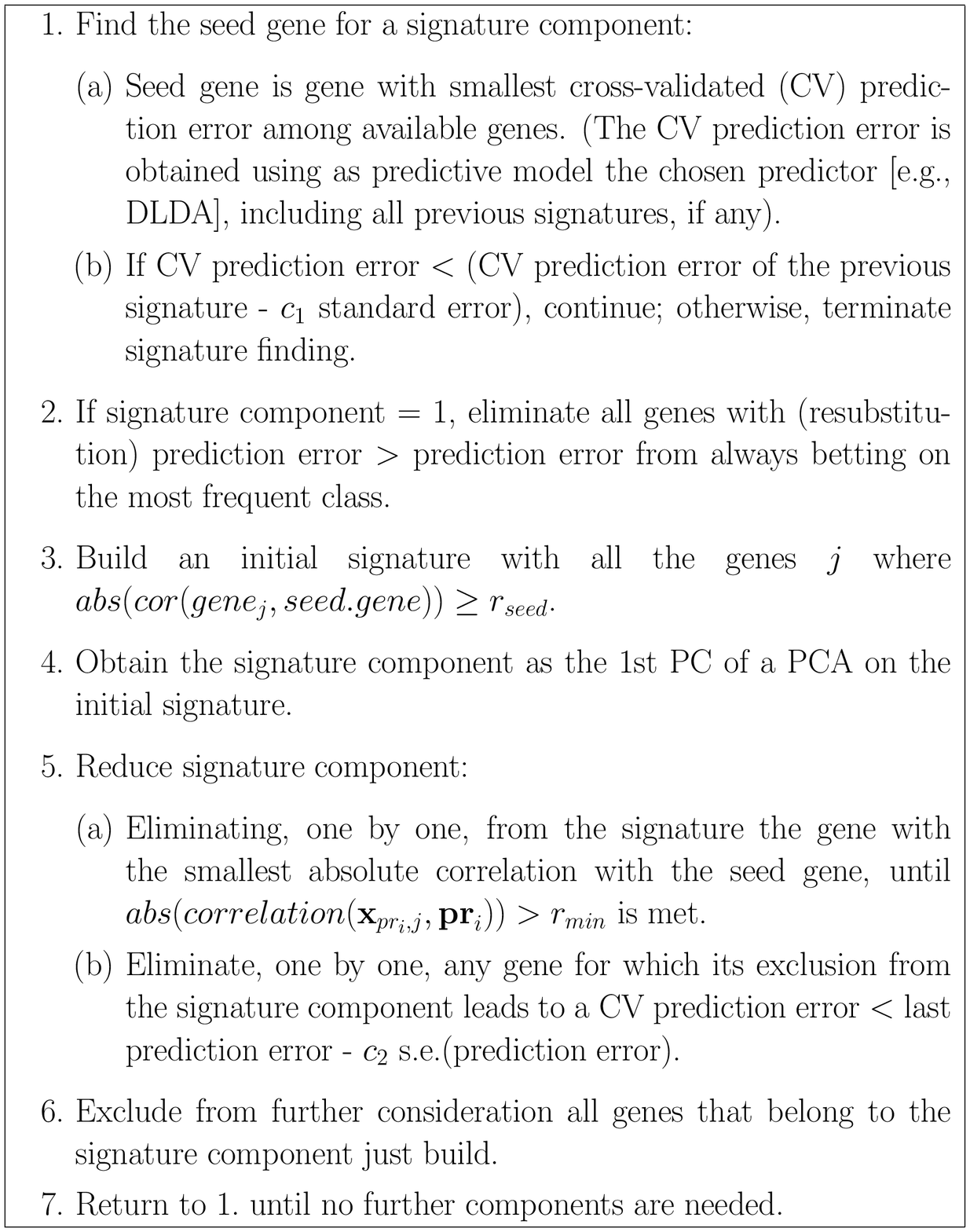}}}
\caption{\label{fig-algorithm} Basic steps of the signature
  algorithm. \hspace{300pt}}
\end{center}
\end{figure}

Predictive performance is evaluated using 10-fold
cross-validation (i.e., the complete algorithm shown in Figure
\ref{fig-algorithm} is applied to each of the 10 ``training sets''). 
This 10-fold cross-validation  was repeated 20 times under each condition.
The error rates shown are not
the CV error rates obtained in steps 1 and 5 of the signature algorithm (see
Figure \ref{fig-algorithm}), since those are biased down (see supplementary
material);
the error rates shown are the error rates obtained from cross-validating the
complete procedure.

Because an important parameter of our method might be $r_{min}$, the minimal
absolute correlation between each gene in a signature component and the
signature component, we have evaluated the performance of the signature method
using a set of values of $r_{min}$ that covers a ``biologically interesting''
range: $\{.60, .65, .70, .75, .80, .85, .90, .95\}$.  In addition, we have also
examined the differences between using $c_1 = c_2 = 1$ compared to $c_1 = c_2 =
0$; the first corresponds to the usual ``1 se rule'' and should lead to more
interpretable results ($c_1$ and $c_2$ are related to how much we penalize
adding a new signature component and how much we penalize eliminating genes
from signature components; see details in supplementary material).

\subsection{The data sets}
\begin{description}
\item[Leukemia dataset] From \cite{golub}. The original data, from an
  Affymetrix chip, comprises 6817 genes, but after filtering as done by the
  authors we are left with 3051 genes. Filtering and preprocessing is described
  in the original paper and in \cite{dudoit-dlda}. We used the training data
  set of 38 cases (27 ALL and 11 AML) in the original paper (the observations
  in the ``test set'' are from a different lab and were collected at different
  times). This data set is available from
  [\url{http://www-genome.wi.mit.edu/cgi-bin/cancer/datasets.cgi}] and also
  from the Bioconductor package multtest ([\url{http://www.bioconductor.org}]).
  
\item[Adenocarcinoma dataset] From \cite{ramas-03}. We used the data from the
  12 metastatic tumors and 64 primary tumors.  The original data set included
  16063 genes from Affymetrix chips.  The data (DatasetA\_Tum\_vsMet.res),
  downloaded from [\url{http://www-genome.wi.mit.edu/cgi-bin/cancer/}], had
  already been rescaled by the authors. We took the subset of 9376 genes
  according to the UniGene mapping, thresholded the data, and filtered by
  variation as explained by the authors.  The final data set contains 9868
  clones (several genes were represented by more than one clone); of these, 196
  had constant values over all individuals.
  
\item[NCI 60 dataset] From \cite{ross}. The data, from cDNA arrays, can be
  obtained from
  [\url{http://genome-www.stanford.edu/sutech/download/nci60/index.html}].  The
  raw data we used, which is the same as the data used in \cite{dudoit-dlda,
    dett-03}, is the one in the file ``figure3.cdt''. As in \cite{dudoit-dlda,
    dett-03} we filtered out genes with more than two missing observations and
  we also eliminated, because of small sample size, the two prostate cell line
  observations and the unknown observation. After filtering, we were left with
  a 61 x 5244 matrix, corresponding to eight different tumor types (note that,
  as done by previous authors, we did not average the two observations with
  triplicate hybridizations). As in \cite{dudoit-dlda} we used 5-nearest
  neighbor imputation of missing data using the program GEPAS \citep{gepas}
  (\url{http://gepas.bioinfo.cnio.es/cgi-bin/preprocess}); unlike
  \cite{dudoit-dlda}, however, we measured gene similarity using Euclidean
  distance from the genes with complete data, instead of correlation:
  \cite{troya} found Euclidean distance to be an appropriate metric.  Finally,
  as in \cite[p.~82]{dudoit-dlda} gene expression data were standardized so
  that arrays had mean 0 and variance 1 across variables (genes).
  
\item[Breast cancer dataset] From \cite{vveer}.  The data, from Affymetrix
  arrays, were downloaded from
  [\url{http://www.rii.com/publications/2002/vantveer.htm}] (we used the files
  ArrayData\_less\_than\_5yr.zip, ArrayData\_greater\_than\_5yr.zip,
  ArrayData\_BRCA1.zip, corresponding to 34 patients that developed distant
  metastases within 5 years, 44 that remained disease-free for over 5 years,
  and 18 with BRCA1 germline mutations and 2 with BRCA2 mutations). As did by
  the authors, we selected only the genes that were ``significantly regulated''
  (see their definition in the paper and supplementary material), which
  resulted in a total of 4869 clones.  Because of the small sample size, we
  excluded the 2 patients with the BRCA2 mutation. We used 5-nearest neighbor
  imputation for the missing data, as for the NCI 60 data set.  Finally, we
  excluded from the analyses the 10th subject from the set that developed
  metastases in less than 5 years (sample 54, IRI000045837, in the original
  data files), because it had 10896 missing values out of the original 24481
  clones, and was an outstanding outlying point both before and after
  imputation.  The breast cancer dataset was used both for two class comparison
  (those that developed metastases within 5 years vs. those that remain
  metastases free after 5 years) and for three group comparisons.
\end{description}

Therefore, we use three datasets in which the problem is classification into
two classes (leukemia, adenocarcinoma, breast cancer), one dataset with a three
class problem (breast cancer) and one dataset with an eight class problem.

\subsection{The competing methods}
We have used three methods that have shown good performance in reviews of
classification methods with microarray data \citep{dudoit-dlda, romualdi-03}.

\begin{description}
\item[Diagonal Linear Discriminant Analysis (DLDA)] DLDA is the maximum
  likelihood discriminant rule, for multivariate normal class densities, when
  the class densities have the same diagonal variance-covariance matrix (i.e.,
  variables are uncorrelated, and for each variable, its variance is the same
  in all classes).  This yields a simple linear rule, where a sample is
  assigned to the class $k$ which minimizes $\Sigma_{j = 1}^p (x_j -
  \bar{x}_{kj})^2/\hat{\sigma}^2_j$, where $p$ is the number of variables, $x_j$ is
  the value on variable (gene) $j$ of the test sample, $\bar{x}_{kj}$ is the
  sample mean of class $k$ and variable (gene) $j$, and $\hat{\sigma}^2_j$ is the
  (pooled) estimate of the variance of gene $j$ \citep{dudoit-dlda}. In spite of
  its simplicity and its somewhat unrealistic assumptions (independent
  multivariate normal class densities), this method has been found to work very
  well.
  
\item[K nearest neighbor (KNN)] KNN is a non-parametric classification method
  that predicts the sample of a test case as the majority vote among the k
  nearest neighbors of the test case \citep{ripley-96, htf-01}.  To decide on
  ``nearest'' here we use, as in \cite{dudoit-dlda}, the Euclidean distance.
  The number of neighbors used (k) is chosen by cross-validation as in
  \cite{dudoit-dlda}: for a given training set, the performance of the KNN for
  values of $k$ in $\{1, 3, 5, \ldots, 21\}$ is determined by cross-validation, and
  the $k$ that produces the smallest error is used.
  
\item[Support Vector Machines (SVM)] SVM are becoming increasingly popular
  classifiers in many areas, including microarrays \citep{furey-00, Lee-Lee,
    ramas-svm}.  SVM (with linear kernel, as used here) try to find an optimal
  separating hyperplane between the classes. When the classes are linearly
  separable, the hyperplane is located so that it has maximal margin (i.e., so
  that there is maximal distance between the hyperplane and the nearest point
  of any of the classes) which should lead to better performance on data not
  yet seen by the SVM. When the data are not separable, there is no separating
  hyperplane; in this case, we still try to maximize the margin but allow some
  classification errors subject to the constraint that the total error
  (distance from the hyperplane in the ``wrong side'') is less than a constant.
  For problems involving more than two classes there are several possible
  approaches; the one used here is the ``one-against-one'' approach, as
  implemented in libsvm \cite{libsvm}. Reviews and introductions to SVM can be
  found in \cite{svm, htf-01}.
\end{description}

For each of these three methods we need to decide which of the genes will be
used to build the predictor. Based on the results of \cite{dudoit-dlda} we have
used the 200 genes with the largest $F$-ratio of between to within groups sums
of squares. \cite{dudoit-dlda} found that, for the methods they considered, 200
genes as predictors tended to perform as well as, or better than, smaller
numbers (30, 40, 50 depending on data set).

We evaluated predictive performance using 10-fold cross-validation;
the results shown are from 20 replications of the 10-fold cv process.  In all
cases, cross-validation includes gene selection \citep{ambroise, simon-03}; in
other words, for the three competing methods and the signature algorithm the
selection of genes is carried out within each of the 10 ``training sets'' of
the cross-validation. Thus, we insure that the subjects for which prediction is
performed have not been used for the gene selection process.

\section{Results}

\begin{figure}[h!]
\begin{center}
{\resizebox{\columnwidth}{!}{%
\includegraphics{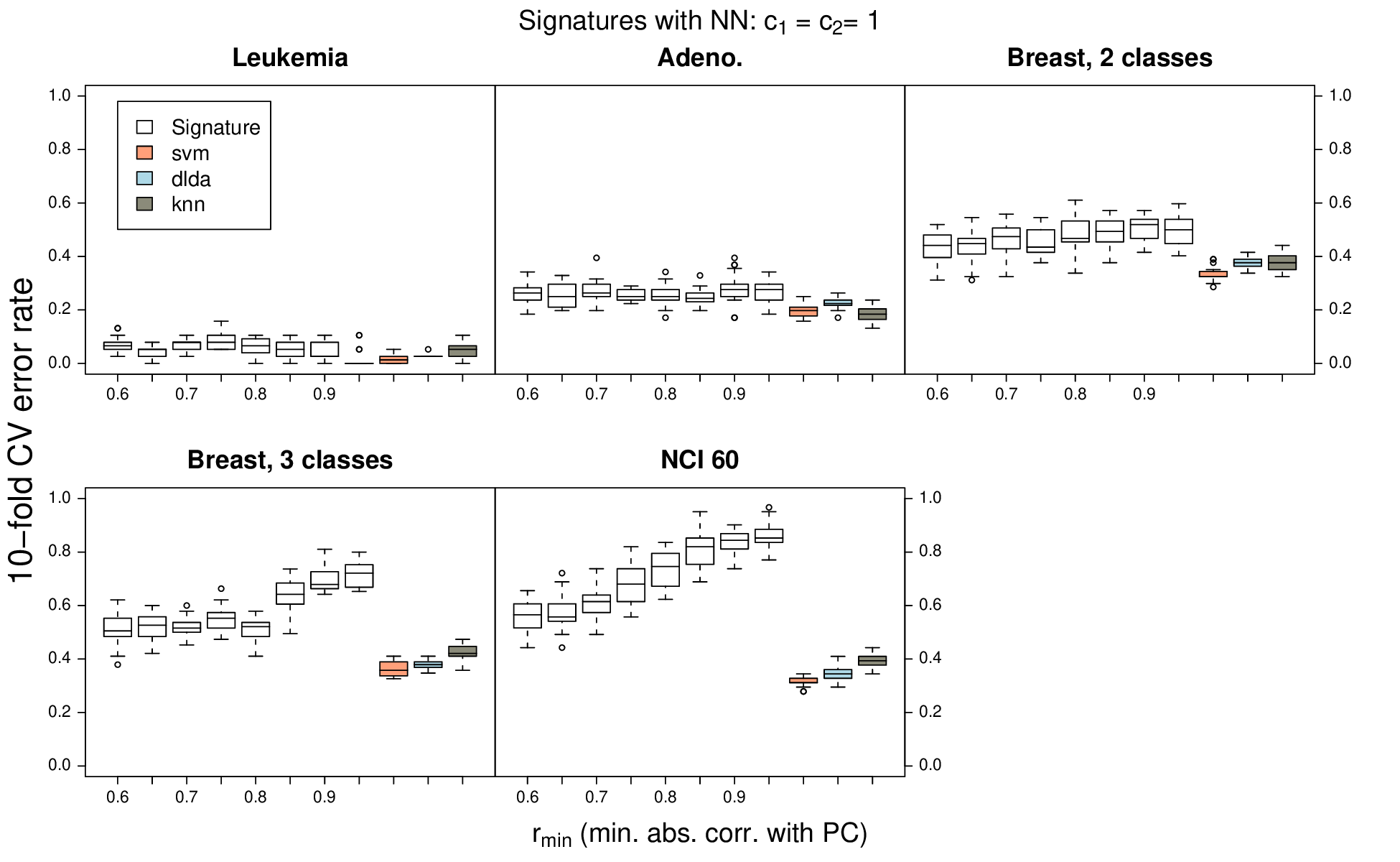}}}
\caption{\label{results-knn}Predictive performance, as a function of
  $r_{min}$, of the signature  method using NN as classifier and comparison with SVM, KNN, and DLDA.
  Figures based on 20 replicates of the 10-fold-CV procedure. Results for $c_1
  = c_2 = 1$.
}
\end{center}
\end{figure}

Predictive performance is shown in Figure \ref{results-knn}. Predictive performance changes
very little from using DLDA vs.\ NN, or setting $c_1 = c_2 = 1$ vs.\ $c_1 = c_2
= 0$; figures for four combinations of classifier and values of $c_1,
c_2$ are shown in the supplementary material.  Comparing $c_1 = c_2 = 1$ with
$c_1 = c_2 = 0$ (see Figures in supplementary material) does not show any
relevant differences in predictive performance; of course, there are differences in the
outcome because, not surprisingly, using $c_1 = c_2 = 0$ tends to result in
more signature components of smaller numbers of genes per component, and higher
correlations between components.  

\vspace{20pt}
\centerline{[Figure 3 about here]}
\vspace{20pt}

\vspace{20pt}
\centerline{[Table 1 about here]}
\vspace{20pt}

With respect to $r_{min}$, except for the
NCI data set, and slightly for the Breast cancer with 3 classes data set,
changes in $r_{min}$ have little effects on predictive performance. In Table
\ref{num.genes} we show the median number of components, median total number of
genes in a signature, and median average number of genes per component obtained in 200
bootstrap runs using $c_1 = c_2 = 1$ with $r_{min} = 0.85$ and $r_{min} = 0.6$,
and using NN as the classifier (see also section \ref{results-repeated}). It
can be seen that changes in $r_{min}$ do affect the outcome in terms
of number of number of genes per component.  These results seem to indicate
that choice of $r_{min}$ can probably be guided more by interpretability
concerns (whether we want larger signature components of looser coexpression or
smaller signature components of tight coexpression) than by concerns over
predictive ability.

Finally, the performance of the signature method is only slightly worse than
that of the three competing classifiers, except for the NCI data set.  As seen
in Table \ref{num.genes}, most of the signatures, specially with
$r_{min} = 0.85$, used very few components of very few genes each (compared to
the use of 200 genes for the competing classifiers); thus, the predictive
performance is achieved using a very small number of genes and thus potentially
facilitating a simple biological interpretation.  In the case of the NCI data
set, there are eight classes with only 61 samples, and in most cases the
signature component only returned between 1 and 3 components. Probably the
forward sequential addition of components in the signature method has affected
negatively the predictive capabilities, because any single addition was most
likely incapable of resulting in a large enough decrease of prediction error to
justify further addition of components (in contrast to using, directly, 200
genes as predictors).

\subsection{\label{results-repeated}Stability of results}

To evaluate the stability of results, we rerun the complete procedure on all
data sets using the bootstrap \citep{efron-tib, boot-dh} with 200 bootstrap
samples, similar to what \cite{efron-gong} do to evaluate a complex fitting
procedure. We run the procedure for settings of $c_1 = c_2 = 1$ with $r_{min} =
0.85$ and $r_{min} = 0.6$, and using NN as the classifier. The results are
shown in Table \ref{table-boot}. The bootstrap results indicate that,
when using a high value of $r_{min}$ we rarely obtain similar solutions
repeatedly; some data sets, however, seem to yield more stable
solutions (e.g., Leukemia data sets) and, not surprisingly, if the $r_{min}$
criterion is set to less stringent values, results tend to be more repeatable.

\vspace{20pt}
\centerline{[Table 2 about here]}
\vspace{20pt}

\section{Discussion}

\subsection{\label{relation-others}Similarities and differences with other methods}

Our method is unique because it simultaneously searches for sets of genes that
are tightly coexpressed and will lead to good predictive performance.  The
search for the sets of genes is carried out using the information from the
dependent variable (at the first stage ---when selecting the seed gene---, and
at the pruning stage of reducing the signature component ---when genes that
lead to decreased predictive performance are eliminated from the signature
component ---; see figure \ref{fig-algorithm}).

One important difference between our proposed method and most previous
approaches that use PCA is that, by performing PCA only on subsets of genes, 
our method returns signature components where genes show tight coexpression.
Because returned components are not orthogonal, and simple components are an
explicit goal, our approach is actually closer to some ideas implemented in
SAS's PROC VARCLUS, which is similar to factor analysis with oblique rotation
and can be used to obtain clusters of variables, of relatively simple
interpretation, to be further used in model building \cite{SAS, varclus}; see
also \cite{harrell-01}.

Using PCA on subsets of genes, instead of the complete set of genes is crucial
because it makes interpretation easier and allows for subsets of tight
coexpression. Simple PCA and related methods \citep{rousson-03, vines,
  jolliffe}
as well as SAS PROC VARCLUS also try to achieve components of tight coexpression, but
many of these approaches cannot be applied with $p \gg n$, and all of them carry
out the PCA without using the information from the dependent variable, which in
our case is a fundamental requirement since the sets of genes with tightest
coexpression would be irrelevant for our purposes if they are not related to
the dependent variable we are trying to model. This difference in objectives is
also evident because our aim is not to explain the most variance in the genes (as in most
simple PCA approaches) nor maximize variable explained across all clusters
(such as in PROC VARCLUS). Overall summarization of information is not
important for our problem, because we are interested in prediction, and
we often have the suspicion that most of the genes in the array are not related
to the outcome variable.  Finally, our approach can result in signature
components which are correlated (sometimes strongly), but this is not
inherently a problem because there is no biological reason to suggest that the
underlying biological causes or factors ought to be independent or
uncorrelated; moreover, if the true underlying causes are not orthogonal, using
a method such as PCA can lead to interpretational and conceptual difficulties
because each biological cause will be spread over several orthogonal components
\citep{interpretation.CPC}: non-orthogonal biological causes are inconsistent
with procedures such as PCA and PLS.

Bayesian classification trees using the 1st PC from gene clusters
\citep{huang-lancet} and block PCA \citep{liu-pca} also used PCA on subsets of
genes, instead of the complete set of genes. In both cases their
subsets of genes were obtained using criteria that did not make any use of the
information from the dependent variable. In addition, in \cite{huang-lancet}
the metagenes are not necessarily of subsets of tightly coexpressed genes. In
\cite{liu-pca} there is an explicit criterion of \% variance accounted for, but
often the number of components used to summarize a subset of genes is too large
to allow for easy interpretation (11 to 16 principal components).

Supervised harvesting of expression genes \citep{harvest} also works with
clusters of subsets of genes, which are then used in a predictive model. As
before, however, the clustering is carried out without using information from
the dependent variable; even if the selection of which subsets or clusters to
use in the model uses the information from the dependent variable, the very
first step of clustering genes does not, and can therefore be unable to recover
sets of tightly coexpressed genes that are good predictors.  Finally, the
``Wilma'' and ``Pelora'' methods \citep{wilma, pelora} do use the information
from the dependent variable in the formation of clusters of genes; however,
there is no explicit objective of achieving tight gene coexpression within
clusters and thus, how tightly coexpressed genes are in each metagene cannot be
specified in advance; in addition, Wilma weights each gene equally within a
cluster (only possible weights are +1 and -1) whereas we use PCA on unscaled
genes, thus allowing genes to play a different role in the specification of the
direction of the signature component (genes with larger among-subject variance
play a more important role in determining direction).

The rest of the alternative methods differ strongly from our proposed method,
either because they do not return subsets of genes but components with loadings
from all genes (e.g., PLS based methods), or return subsets of genes where
there is not requirement of tight coexpression \citep[e.g., weighted gene voting,][]
{golub}.

After gene selection and dimension reduction (i.e., the use of only the 1st PC,
that collapses all the information from the genes of a signature component onto
one dimension), the predictive model of our choice is fitted. In this sense,
the method as presented here is ``just'' a DLDA or NN that uses signature
components instead of genes as the predictors. The choice of DLDA and NN was
made based on published results that showed their excellent performance with
microarray data. In particular, \cite{dudoit-dlda} showed that other forms of
discriminant analysis tended to perform much worse because of the small ratio
sample size/parameters needed to estimate covariances and different variances
per group; however, since in most cases our method returns just a few signature
components, other types of discriminant analysis that use the information from
the covariance of the predictors (e.g., linear discriminant analysis and
quadratic discriminant analysis) might prove useful.

\subsection{\label{within.among}Coexpression: across-group and within-group}

The algorithm can include in a signature component genes that show no
correlation within groups but that show correlation among groups because they
are far apart in the multidimensional space, and the correlation coefficient is
computed across the whole, pooled, sample. The algorithm might even include in
the same signature component genes that have very different patterns of
correlation in different groups, if they still show sufficiently strong
correlation over the pooled sample.
    
To our knowledge, this issue has not been explicitly addressed in any other 
approach to the signature problem (see reviews in supplementary
material).
However, probably the most biologically relevant components are those where
there is strong correlation within groups, because this is a more reliable
indicator of coordinated expression.

A possible solution is, for example, to only accept results for a signature
component if a principal component analysis over the pooled sample after
centering the data with respect to the group means yields a relevant first
eigenvalue; for added robustness, we might want to use the trimmed mean. This
approach, however, does not directly address if there are different
multivariate orientations in different groups of subjects, and how these
orientations within-group relate to the across-group orientation. In
particular, the case where several groups not only have the same first
principal component, but are lined up along a common axis, known as
``allometric extension'' \citep{hills.1982, allometric.extension}, might
constitute the most natural type of signature component.

Krzanowski \citep[see reviews and summary in][]{krza-book, jolliffe} has
proposed a method to directly compare the subspaces defined by the principal
components of each of the groups. In our case, as we only use the first
principal component of a set of genes to define a signature component, for each
signature component we can compare the first principal component of each group.
An example using the NCI 60 data set is shown in the supplementary material.
This method only compares the orientation of the principal components (the
eigenvectors) but does not compare the location of the multivariate means.  The
EDDA \citep{EDDA} and common principal components \citep{flury-88} approaches
provide frameworks to examine differences among covariance matrices of
particular interest in the context of discrimination among groups.
Nevertheless, it must be remembered that biological interpretation of results
is not always straightforward, specially when the underlying biological factors
are not orthogonal \citep{interpretation.CPC}, and that we need to asses the
need for robust methods \citep{robust.CPC} and power issues related to the
small sample sizes common in microarray data. We are currently investigating
some of these issues; however, in the presence of the difficulties associated
with interpretation, low statistical power, and possible lack of robustness to
outlying observations, right now the best recommendation could be to conduct an
examination of patterns of within vs.  among coexpression for the returned
signatures.

\subsection{Stability and biological relevance of signatures}

Our method follows from an operationalization of signatures and signature
components (figure \ref{box-signature}). If the ideas embodied in figure
\ref{box-signature} have empirical support, then it might be possible to build
good predictive models using just a few, very specific biological features that
can be related to alterations on particular pathway. Our method is unique in
this regard because it returns only a few signature components where genes
within each component are tightly coexpressed.


On the other hand, our method can indicate that the data are inconsistent with
the ideas behind figure \ref{box-signature}: if the signature components show
high instability, this is evidence that very different models can be obtained
from the data.  Widely different models from a reasonably large data set cast
doubt on the idea that a few, easily interpretable, signature components
strongly correlated with the expression of a few key genes are associated to
the clinical outcome of interest.  In the context of building predictors from
gene expression data, \cite{Somorjai2003} \citep[see also][]{zhang} have
emphasized that this non-uniqueness leads to interpretational difficulties and
should make researchers skeptical about the biological relevance of any set of
predictors; moreover, they explain how this non-uniqueness can arise from
dataset sparsity.  Of course, both of these issues are relevant to the present
proposed methodology. None of the data sets examined in this paper yield
stable signatures when we use stringent criteria of gene
coexpression (see \ref{results-repeated} and Table \ref{table-boot} with
$r_{min} = 0.85$), although some of the data sets are somewhat stable from run
to run when the $r_{min}$ is set to small values (but other data sets show
signatures that vary widely from run to run).  These results add to the above
references in the sense that biological interpretation should be carried out
very cautiously, and emphasize the difference between attempting to build good
predictors and attempting interpretation \citep[see
also][]{breiman-2-cultures}. More relevant to the current work, the present
results indicate that simple models of molecular signatures warrant further
critical scrutiny, and that it might be extremely hard to identify molecular
signatures from such sparse data sets \citep[see][for a meta-analysis attempt
to identify stable ``signatures'']{meta-signature}.

We must recognize that these results are preliminary for two main reasons.
First, establishing that two or more signature components are different
probably requires additional information besides the identities of the genes;
for instance, information from Gene Ontology, or known participation on certain
regulatory networks. Second, we have used two different $r_{min}$ thresholds,
but it is unclear what constitute ``biologically reasonable'' patterns of
covariation between genes that are to belong to the same signature component.
Nevertheless,  even if preliminary, these results emphasize the need for
further work in the operationalization and explicit definition of what we mean
by molecular signatures, careful consideration of the stability of results, and
critical assessment of the sample sizes need to reliably identify molecular signatures.

\subsection{Alternative statistical methods for alternative biological models}

As mentioned in the introduction, we started by trying to clarify,
conceptually, what is often understood by molecular signature (see Figure
\ref{box-signature}), and then devised a statistical method to fulfill those
requirements. The biological model underlying the suggested method is
one where most of the genes are not relevant for prediction, relevant genes are
involved in one and only one signature component (i.e., non-overlapping
signature components), and the signature components are common, and have
similar covariance matrices, in different groups.

However, other biological models are plausible, and for those biological models
other statistical methods would be more appropriate.  The simultaneous
clustering and classification approach in \cite{Jornsten.Yu2003} could be
extended by placing restrictions on the covariance matrix (i.e., require a
minimum correlation between genes) but possibly allowing for different
covariance matrices among groups; thus, we could address directly issues of
different across vs. within-group correlations (see section
\ref{within.among}), within a formal inferential framework. Biological models
where signature components are not common and/or do not behave similarly in
different groups could be investigated using modifications of the Plaid model
of \cite{plaid} \citep[see also][]{plaid-heather}.  In addition, genes with the
highest correlation need not be the best candidates for being in the same
biological pathway; activity in a pathway might just require that precursor
genes get activated, but once a threshold is reached, it might not be very
important by how much the threshold is exceeded. This type of behavior could
preclude strong correlation between genes that belong to the same pathway.
This can be modelled building upon the latent class methods of Parmigiani and
colleagues \citep{poe1, parmigiani-poe2, Scharpf.Parmigiani2003}, where
signature components are based on under-, over- or baseline expression (instead
of expression levels). Work along these lines is currently in progress in our
group.

\subsection{``Just'' dimensional reduction?}

Even if the current method fails to identify stable features that can be
associated to molecular signatures, it can be a useful dimension reduction
tool. Difficulties associated with a simple mapping of the returned ``signature
components'' to pathways, and problems derived from instability of the found
components also affect any other of the existing alternative methods. Thus, in
the presence of instability of results, it is more appropriate to regard this
method as dimension reduction tool that could lead to simple biological
interpretation.  The simple biological interpretation could be helped not only
because of the coexpression of the genes that make a signature component, but
also because the dimension reduction performed is quite remarkable compared to
other methods (the number of signature components and genes returned is very
small for the five data sets examined; see Results) with, at most, only a
slight decrease in predictive performance. Moreover, the user can control the
relative trade-offs between predictive performance and potential
interpretability of results (e.g., coexpression of sets of genes) by changing
the $r_{min}$ parameter (note that if $r_{min} = 1$ the method becomes
essentially either DLDA or NN with forward addition of genes to the model).
This flexible modification of the trade-offs between prediction error and
interpretability is of great importance in methods that are largely exploratory
and oriented towards providing ``biologically interpretable'' output; in other
words, methods for which minimization of prediction error should not be the
only goal.

\section{Conclusions}

The most common methods for finding signatures present several deficiencies
that do not allow them to return signatures and signature components that
fulfill basic biological requirements. After suggesting an operational
definition of signature and signature components, we have developed a method
that follows directly from what are often considered as the biologically
relevant signature characteristics. The method developed returns signature
components of tightly coexpressed genes and thus can facilitate biological
interpretation.  In this paper we have applied the method to classification
problems, but this approach in fact sets up a framework that allows us to find
signatures regardless of the type of dependent variable. Extension to use other
classifiers is straightforward and it should also be easy to incorporate other
types of dependent variables to allow, for example, survival analysis.  We have
also shown that the predictive performance of our method is comparable to that
of state of the art methods. Finally, our method not only could facilitate
mapping pathological alterations to a few, tightly coexpressed sets of genes,
but can also provide evidence that the underlying biological assumptions behind
this attempt are inconsistent with the data. In the five data sets analyzed,
our results suggest that identification of molecular signatures is questionable
under this simple model.  These results emphasize the needed for further work
on the operationalization of the biological model and the necessity of critical
assessment of the stability of putative signatures.

\section{Acknowledgements}

S.\ Ramaswamy for answering questions about the data and processing steps for
the adenocarcinoma dataset, H.\ Dai for explanations about the breast cancer
dataset, and S.\ Dudoit for answering some questions about the NCI data set.
A.\ Pérez and M.\ A.\ Piris for introducing me to, and emphasizing the
importance of, molecular signatures. The Bioinformatics Unit at CNIO, C.\ 
Lázaro-Perea, Y.~Benjamini and F.~Falciani for discussion. D.\ Casado for help
with the code. C.\ Lázaro-Perea and two anonymous reviewers for detailed and
thorough comments on the ms. The author was partially supported by the Ramón y
Cajal program of the Spanish MCyT (Ministry of Science and Technology); funding
partially provided by project TIC2003-09331-C02-02 of the Spanish MCyT.

\bibliography{signatures}

\begin{thebibliography}{}

\bibitem[Alizadeh {\em et~al.}, 2000]{alizadeh}
Alizadeh, A.~A., Eisen, M.~B., Davis, R.~E., Ma, C., Lossos, I.~S., Rosenwald,
  A., Boldrick, J.~C., Sabet, H., Tran, T., Yu, X., Powell, J.~I., Yang, L.,
  Marti, G.~E., Moore, T., Hudson, J.~J., Lu, L., Lewis, D.~B., Tibshirani, R.,
  Sherlock, G., Chan, W.~C., Greiner, T.~C., Weisenburger, D.~D., Armitage,
  J.~O., Warnke, R.  \& Staudt, L.~M. (2000{\em{}}) Distinct types of diffuse
  large b-cell lymphoma identified by gene expression profiling.
\newblock {\em Nature, } {\bf 403}, 503--511.

\bibitem[Ambroise \& McLachlan, 2002]{ambroise}
Ambroise, C. \& McLachlan, G.~J. (2002{\em{}}) Selection bias in gene
  extraction on the basis of microarray gene-expression data.
\newblock {\em Proc Natl Acad Sci USA, } {\bf 99} (10), 6562--6566.

\bibitem[Antoniadis {\em et~al.}, 2003]{mave-dna}
Antoniadis, A., Lambert-Lacroix, S.  \& Leblanc, F. (2003{\em{}}) Effective
  dimension reduction methods for tumor classification using gene expression
  data.
\newblock {\em Bioinformatics, } {\bf 19} (5), 563--570.

\bibitem[Bartoletti {\em et~al.}, 1999]{allometric.extension}
Bartoletti, S., Flury, B.~D.  \& Nel, D.~G. (1999{\em{}}) Allometric extension.
\newblock {\em Biometrics, } {\bf 55}, 1210--1214.

\bibitem[Bensmail \& Celeux, 1996]{EDDA}
Bensmail, H. \& Celeux, G. (1996{\em{}}) Regularized gaussian discriminant
  analysis through eigenvalue decomposition.
\newblock {\em Journal American Statistical Association, } {\bf 91},
  1743--1748.

\bibitem[Boente {\em et~al.}, 2002]{robust.CPC}
Boente, G., Pires, A.~M.  \& Rodrigues, I. (2002{\em{}}) Influence functions
  and outlier detection under the common principal components model: a robust
  approach.
\newblock {\em Biometrika, } {\bf 89}, 861--875.

\bibitem[Breiman, 2001]{breiman-2-cultures}
Breiman, L. (2001{\em{}}) Statistical modeling: the two cultures (with
  discussion).
\newblock {\em Statistical Science, } {\bf 16}, 199--231.

\bibitem[Burgues, 1998]{svm}
Burgues, C. J.~C. (1998{\em{}}) A tutorial on support vector machines for
  pattern recognition.
\newblock {\em Knowledge Discovery and Data Mining, } {\bf 2}, 121--167.

\bibitem[Chang \& Lin, 2003]{libsvm}
Chang, C.-C. \& Lin, C.-J. (2003{\em{}}).
\newblock Libsvm: a library for support vector machines.
\newblock Technical report Department of Computer Science, National Taiwan
  University URL: http://www.csie.ntu.edu.tw/~cjlin/libsvm.

\bibitem[Davison \& Hinkley, 1997]{boot-dh}
Davison, A.~C. \& Hinkley, D.~V. (1997{\em{}}) {\em Bootstrap methods and their
  application}.
\newblock Cambridge University Press, Cambridge.

\bibitem[Dettling \& B{\"u}hlmann, 2002]{wilma}
Dettling, M. \& B{\"u}hlmann, P. (2002{\em{}}) Supervised clustering of genes.
\newblock {\em Genome Biology, } {\bf 3} (12), 0069.1--0069.15.

\bibitem[Dettling \& B{\"u}hlmann, 2003]{dett-03}
Dettling, M. \& B{\"u}hlmann, P. (2003{\em{}}) Boosting for tumor
  classification with gene expression data.
\newblock {\em Bioinformatics, } {\bf 19} (9), 1061--1069.

\bibitem[Dettling \& B{\"u}hlmann, 2004]{pelora}
Dettling, M. \& B{\"u}hlmann, P. (2004{\em{}}) Finding predictive gene groups
  from microarray data.
\newblock {\em J. Multivariate Anal., } {\bf 90}, 106--131.

\bibitem[Dudoit {\em et~al.}, 2002]{dudoit-dlda}
Dudoit, S., Fridlyand, J.  \& Speed, T.~P. (2002{\em{}}) Comparison of
  discrimination methods for the classification of tumors suing gene expression
  data.
\newblock {\em J Am Stat Assoc, } {\bf 97} (457), 77--87.

\bibitem[Efron \& Gong, 1983]{efron-gong}
Efron, B. \& Gong, G. (1983{\em{}}) A leisurely look at the bootstrap, the
  jacknife, and cross-validation.
\newblock {\em Am Stat, } {\bf 37} (1), 36--48.

\bibitem[Efron \& Tibshirani, 1993]{efron-tib}
Efron, B. \& Tibshirani, R.~J. (1993{\em{}}) {\em An introduction to the
  bootstrap}.
\newblock Chapman and Hall, London.

\bibitem[Flury, 1988]{flury-88}
Flury, B. (1988{\em{}}) {\em Common principal components and related
  techniques}.
\newblock John Wiley \& Sons, New York.

\bibitem[Furey {\em et~al.}, 2000]{furey-00}
Furey, T.~S., Cristianini, N., Duffy, N., Bednarski, D.~W., Schummer, M.  \&
  Haussler, D. (2000{\em{}}) Support vector machine classification and
  validation of cancer tissue samples using microarray expression data.
\newblock {\em Bioinformatics, } {\bf 16} (10), 906--914.

\bibitem[Garrett \& Parmigiani, 2003]{parmigiani-poe2}
Garrett, E. \& Parmigiani, G. (2003{\em{}}) {\em The analysis of gene
  expression data: methods and software}. New York: Springer pp. 362--387.

\bibitem[Golub {\em et~al.}, 1999]{golub}
Golub, T.~R., Slonim, D.~K., Tamayo, P., Huard, C., Gaasenbeek, M., Mesirov,
  J.~P., Coller, H., Loh, M.~L., Downing, J.~R., Caligiuri, M.~A., Bloomfield,
  C.~D.  \& Lander, E.~S. (1999{\em{}}) Molecular classification of cancer:
  class discovery and class prediction by gene expression monitoring.
\newblock {\em Science, } {\bf 286}, 531--537.

\bibitem[Harrell, 2001]{harrell-01}
Harrell, J. F.~E. (2001{\em{}}) {\em Regression modeling strategies}.
\newblock Springer, New York.

\bibitem[Hastie {\em et~al.}, 2001{\em{a}}]{harvest}
Hastie, T., Tibshirani, R., Botstein, D.  \& Brown, P. (2001{\em{a}})
  Supervised harvesting of expression trees.
\newblock {\em Genome Biology, } {\bf 2}, 0003.1--0003.12.

\bibitem[Hastie {\em et~al.}, 2001{\em{b}}]{htf-01}
Hastie, T., Tibshirani, R.  \& Friedman, J. (2001{\em{b}}) {\em The elements of
  statistical learning}.
\newblock Springer, New York.

\bibitem[Hedenfalk {\em et~al.}, 2001]{hedenf}
Hedenfalk, I., Duggan, D., Chen, Y., Radmacher, M., Bittner, M., Simon, R.,
  Meltzer, P., Gusterson, B., Esteller, M., Kallioniemi, O., Wilfond, B., Borg,
  A.  \& Trent, J. (2001{\em{}}) Gene-expression profiles in hereditary breast
  cancer.
\newblock {\em N Engl J Med, } {\bf 344} (8), 539--548.

\bibitem[Herrero {\em et~al.}, 2003]{gepas}
Herrero, J., D{\'\i}az-Uriarte, R.  \& Dopazo, J. (2003{\em{}}) Gene expression
  data preprocessing.
\newblock {\em Bioinformatics, } {\bf 19} (5), 655--656.

\bibitem[Hills, 1982]{hills.1982}
Hills, M. (1982{\em{}}) {\em Encyclopedia of statistical sciences, Volume I}.
  New York: Wiley pp. 48--54.

\bibitem[Houle {\em et~al.}, 2002]{interpretation.CPC}
Houle, D., Mezey, J.  \& Galpern, P. (2002{\em{}}) Interpretation of the
  results of common principal components analyses.
\newblock {\em Evolution, } {\bf 56}, 433--440.

\bibitem[Huang {\em et~al.}, 2003{\em{a}}]{huang-lancet}
Huang, E., Cheng, S.~H., Dressman, H., Pittman, J., Tsou, M.-H., Horng, C.-F.,
  Bild, A., Iversen, E.~S., Liao, M., Chen, C.-M., West, M., Nevins, J.~R.  \&
  Huang, A.~T. (2003{\em{a}}) Gene expression predictors of breast cancer
  outcomes.
\newblock {\em Lancet, } {\bf 361}, 1590--1596.

\bibitem[Huang {\em et~al.}, 2003{\em{b}}]{huang-nature}
Huang, E., Ishida, S., Pittman, J., Dressman, H., Bild, A., Kloos, M., D'Amico,
  M., Pestell, R.~G., West, M.  \& Nevins, J.~R. (2003{\em{b}}) Gene expression
  phenotypic models that predict the activity of oncogenic pathways.
\newblock {\em Nature Genetics, } {\bf 34} (2), 226--230.

\bibitem[Jolliffe, 2002]{jolliffe}
Jolliffe, I.~T. (2002{\em{}}) {\em Principal component analysis, 2nd ed.}
\newblock Springer, New York.

\bibitem[J{\"o}rnsten \& Yu, 2003]{Jornsten.Yu2003}
J{\"o}rnsten, R. \& Yu, B. (2003{\em{}}) {Simultaneous gene clustering and
  subset selection for sample classification via MDL}.
\newblock {\em Bioinformatics, } {\bf 19}, 1100--1109.

\bibitem[Krzanowski, 1998]{krza-book}
Krzanowski, W.~J. (1998{\em{}}) {\em Principles of multivariate analysis}.
\newblock Oxford University Press, Oxford.

\bibitem[Lazzeroni \& Owen, 2002]{plaid}
Lazzeroni, L. \& Owen, A. (2002{\em{}}) Plaid models for gene expression data.
\newblock {\em Statistica Sinica, } {\bf 12}, 61--86.

\bibitem[Lee \& Lee, 2003]{Lee-Lee}
Lee, Y. \& Lee, C.-K. (2003{\em{}}) Classification of multiple cancer types by
  multicategory support vector machines using gene expression data.
\newblock {\em Bioinformatics, } {\bf 19} (9), 1132--1139.

\bibitem[Liu {\em et~al.}, 2002]{liu-pca}
Liu, A., Zhang, Y., Gehan, E.  \& Clarke, R. (2002{\em{}}) Block principal
  component analysis with application to gene microarray data classification.
\newblock {\em Statist Med, } {\bf 21}, 3465--3474.

\bibitem[Morrison, 1990]{morrison}
Morrison, D.~F. (1990{\em{}}) {\em Multivariate statistical methods}.
\newblock McGraw-Hill, New York.

\bibitem[Nelson, 2001]{varclus}
Nelson, B.~D. (2001{\em{}}) Variable reduction for modeling using proc varclus.
\newblock In {\em Proceedings of the Twenty-Sixth Annual SAS Users Group
  International Conference}, (Institute, S., ed.), SAS Institute, Cary, NC.

\bibitem[Parmigiani {\em et~al.}, 2002]{poe1}
Parmigiani, G., Garrett, E., Anbazhaghan, R.  \& Gabrielson, E. (2002{\em{}}) A
  statistical framework for expression-based molecular classification in
  cancer.
\newblock {\em J. Royal Statistical Society, Series B, } {\bf 64}, 717--736.

\bibitem[Pomeroy {\em et~al.}, 2002]{pomeroy}
Pomeroy, S., Tamayo, P., Gaasenbeek, M., Sturla, L., Angelo, M., McLaughlin,
  M., Kim, J., Goumnerova, L., Black, P., Lau, C., Allen, J., Zagzag, D.,
  Olson, J., Curran, T., Wetmore, C., Biegel, J., Poggio, T., Mukherjee, S.,
  Rifkin, R., Califano, A., Stolovitzky, G., Louis, D., Mesirov, J., Lander, E.
   \& Golub, T. (2002{\em{}}) Prediction of central nervous system embryonal
  tumour outcome based on gene expression.
\newblock {\em Nature, } {\bf 415}, 436--442.

\bibitem[Ramaswamy {\em et~al.}, 2003]{ramas-03}
Ramaswamy, S., Ross, K.~N., Lander, E.~S.  \& Golub, T.~R. (2003{\em{}}) A
  molecular signature of metastasis in primary solid tumors.
\newblock {\em Nature Genetics, } {\bf 33}, 49--54.

\bibitem[Ramaswamy {\em et~al.}, 2001]{ramas-svm}
Ramaswamy, S., Tamayo, P., Rifkin, R., Mukherjee, S., Yeang, C., Angelo, M.,
  Ladd, C., Reich, M., Latulippe, E., Mesirov, J., Poggio, T., Gerald, W.,
  Loda, M., Lander, E.  \& Golub, T. (2001{\em{}}) Multiclass cancer diagnosis
  using tumor gene expression signatures.
\newblock {\em Proc Natl Acad Sci USA, } {\bf 98} (26), 15149--15154.

\bibitem[Rhodes {\em et~al.}, 2004]{meta-signature}
Rhodes, D., Yu, J., Shanker, K., Deshpande, N., Varambally, R., Ghosh, D.,
  Barrette, T., Pandey, A.  \& Chinnaiyan, A. (2004{\em{}}) Large-scale
  meta-analysis of cancer microarray data identifies common transcriptional
  profiles of neoplastic transformation and progression.
\newblock {\em PNAS, } {\bf 101}, 9309--9314.

\bibitem[Ripley, 1996]{ripley-96}
Ripley, B.~D. (1996{\em{}}) {\em Pattern recognition and neural networks}.
\newblock Cambridge University Press, Cambridge.

\bibitem[Romualdi {\em et~al.}, 2003]{romualdi-03}
Romualdi, C., Campanaro, S., Campagna, D., Celegato, B., Cannata, N., Toppo,
  S., Valle, G.  \& Lanfranchi, G. (2003{\em{}}) Pattern recognition in gene
  expression profiling using dna array: a comparative study of different
  statistical methods applied to cancer classification.
\newblock {\em Hum. Mol. Genet., } {\bf 12} (8), 823--836.

\bibitem[Rosenwald {\em et~al.}, 2002]{rosenwald}
Rosenwald, A., Wright, G., Chan, W.~C., Connors, J.~M., Campo, E., Fisher,
  R.~I., Gascoyne, R.~D., Muller-Hermelink, H.~K., Smeland, E.~B., Giltnane,
  J.~M., Hurt, E.~M., Zhao, H., Averett, L., Yang, L., Wilson, W.~H., Jaffe,
  E.~S., Simon, R., Klausner, R.~D., Powell, J., Duffey, P.~L., Longo, D.~L.,
  Greiner, T.~C., Weisenburger, D.~D., Sanger, W.~G., Dave, B.~J., Lynch,
  J.~C., Vose, J., Armitage, J.~O., Montserrat, E., Lopez-Guillermo, A.,
  Grogan, T.~M., Miller, T.~P., LeBlanc, M., Ott, G., Kvaloy, S., Delabie, J.,
  Holte, H., Krajci, P., Stokke, T., Staudt, L.~M.  \& the Lymphoma/Leukemia
  Molecular Profiling~Project (2002{\em{}}) The use of molecular profiling to
  predict survival after chemotherapy for diffuse large-b-cell lymphoma.
\newblock {\em N Engl J Med, } {\bf 346} (25), 1937--1947.

\bibitem[Ross {\em et~al.}, 2000]{ross}
Ross, D.~T., Scherf, U., Eisen, M.~B., Perou, C.~M., Rees, C., Spellman, P.,
  Iyer, V., Jeffrey, S.~S., de~Rijn, M.~V., Waltham, M., Pergamenschikov, A.,
  Lee, J.~C., Lashkari, D., Shalon, D., Myers, T.~G., Weinstein, J.~N.,
  Botstein, D.  \& Brown, P.~O. (2000{\em{}}) Systematic variation in gene
  expression patterns in human cancer cell lines.
\newblock {\em Nature Genetics, } {\bf 24} (3), 227--235.

\bibitem[Rousson \& Gasser, 2003]{rousson-03}
Rousson, V. \& Gasser, T. (2003{\em{}}).
\newblock Simple component analysis.
\newblock Technical report Department of Biostatistics, University of Z?rich,
  Switzerland.

\bibitem[SAS~Insitute, 1999]{SAS}
SAS~Insitute, I. (1999{\em{}}) {\em SAS/STAT User's guide}.
\newblock SAS Institute Inc, Cary, NC.

\bibitem[Scharpf {\em et~al.}, 2003]{Scharpf.Parmigiani2003}
Scharpf, R., Garrett, E.~S., Hu, J.  \& Parmigiani, G. (2003{\em{}})
  {Statistical modeling and visualization of molecular profiles in cancer}.
\newblock {\em Biotechniques, } {\bf Suppl}, 22--29.

\bibitem[Shaffer {\em et~al.}, 2001]{shaffer}
Shaffer, A., Rosenwald, A., Hurt, E., Giltnane, J., Lam, L., Pickeral, O.  \&
  Staudt, L. (2001{\em{}}) Signatures of the immune response.
\newblock {\em Immunity, } {\bf 15}, 375--385.

\bibitem[Shipp {\em et~al.}, 2002]{shipp}
Shipp, M.~A., Ross, K.~N., Tamayo, P., Weng, A.~P., Kutok, J.~L., Aguiar, R.
  C.~T., Gaasenbeek, M., Angelo, M., Reich, M., Pinkus, G.~S., Ray, T.~S.,
  Koval, M.~A., Last, K.~W., Norton, A., Lister, T.~A., Mesirov, J., Neuberg,
  D.~S., Lander, E.~S., Aster, J.~C.  \& Golub, T.~R. (2002{\em{}}) Diffuse
  large b-cell lymphoma outcome prediction by gene-expression profiling and
  supervised machine learning.
\newblock {\em Nature Medicine, } {\bf 8} (1), 68--74.

\bibitem[Simon {\em et~al.}, 2003]{simon-03}
Simon, R., Radmacher, M.~D., Dobbin, K.  \& McShane, L.~M. (2003{\em{}})
  Pitfalls in the use of dna microarray data for diagnostic and prognostic
  classification.
\newblock {\em Journal of the National Cancer Institute, } {\bf 95} (1),
  14--18.

\bibitem[Somorjai {\em et~al.}, 2003]{Somorjai2003}
Somorjai, R.~L., Dolenko, B.  \& Baumgartner, R. (2003{\em{}}) {Class
  prediction and discovery using gene microarray and proteomics mass
  spectroscopy data: curses, caveats, cautions}.
\newblock {\em Bioinformatics, } {\bf 19}, 1484--1491.

\bibitem[Troyanskaya {\em et~al.}, 2001]{troya}
Troyanskaya, O., Cantor, M., Sherlock, G., Brown, P., Hastie, T., Tibshirani,
  R., Botstein, D.  \& Altman, R. (2001{\em{}}) Missing value estimation
  methods for dna microarrays.
\newblock {\em Bioinformatics, } {\bf 17}, 520--525.

\bibitem[Turner {\em et~al.}, 2004]{plaid-heather}
Turner, H., Bailey, T.  \& Krzanowski, W. (2004{\em{}}) Improved biclustering
  of microarray data demonstrated through systematic performance tests.
\newblock {\em Comput. Statist. Data Anal., } {\bf }, In press.

\bibitem[van Belle, 2002]{vveer}
van Belle, G. (2002{\em{}}) Gene expression profiling predicts clinical outcome
  of breast cancer.
\newblock {\em Nature, } {\bf 415}, 530--536.

\bibitem[Vines, 2000]{vines}
Vines, S.~K. (2000{\em{}}) Simple principal components.
\newblock {\em Applied Statistics, } {\bf 49} (4), 441--451.

\bibitem[West {\em et~al.}, 2001]{West-pnas}
West, M., Blanchette, C., Dressman, H., Huang, E., Ishida, S., Spang, R.,
  Zuzan, H., Olson, J. A.~J., Marks, J.~R.  \& Nevins, J.~R. (2001{\em{}})
  Predicting the clinical status of human breast cancer by using gene
  expression profiles.
\newblock {\em Proc Natl Acad Sci USA, } {\bf 98} (20), 11462--11467.

\bibitem[Zhang {\em et~al.}, 2003]{zhang}
Zhang, H., Yu, C.-Y.  \& Singer, B. (2003{\em{}}) Cell and tumor classification
  using gene expression data: construction of forests.
\newblock {\em Proc Natl Acad Sci USA, } {\bf 100} (7), 4168--4172.

\end{thebibliography}
\bibliographystyle{bioinformatics}


\newpage

\begin{table}[ph!]
\begin{center}
\begin{tabular}{l|p{0.85cm}p{1.75cm}p{2.75cm}|p{0.85cm}p{1.75cm}p{2.75cm}}
& \multicolumn{3}{c}{$r_{min} = 0.85$} \vline & \multicolumn{3}{c}{$r_{min} = 0.6$}\\
\cline{2-7}
Data set & Total genes & \# Components & Mean Genes/Component & Total genes & \# Components & Mean Genes/Component \\
\hline
Leukemia & \ 6 & \ 1 & \ 5 & \ 52.5 & \ 1 & \ 50 \\
Breast cancer (2 classes) &  \ 2 & \ 2 & \ 1 & \ 10 & \ 2 & \ \ 3.875 \\
Breast cancer (3 classes) &  \ 6 & \ 2 & \ 2.5 & \ 63.5 & \ 2 & \ 33.25\\
Adenocarcinoma &  \ 5 & \ 1 & \ 3 & \ 45.5 & \ 1 & \ 31.5\\
NCI 60 &  \ 4 & \ 2 & \ 1.67 & \ 16 & \ 2 & \ \ 7.1 \\
\hline
\end{tabular}

\caption{\label{num.genes}Median values from 200 bootstrap runs for total
  number of genes in signatures, number of signature components and average
  number of genes per component.}

\end{center}
\end{table}

\vspace{70pt}
\newpage

\begin{table}[ph!]
\begin{center}
\begin{tabular}{l|p{0.45cm}p{0.45cm}p{0.45cm}|p{0.45cm}p{0.45cm}p{0.45cm}}
& \multicolumn{3}{c}{$r_{min} = 0.85$} \vline & \multicolumn{3}{c}{$r_{min} = 0.6$}\\
\cline{2-7}
Data set& \multicolumn{3}{c}{Genes present} \vline & \multicolumn{3}{c}{Genes present}\\
& \multicolumn{3}{c}{at least \% runs} \vline & \multicolumn{3}{c}{at least \% runs}\\
& 50 & 20 & 10 & 50 & 20 & 10 \\
\hline
Leukemia & \  0 &  \ 7 &15 & 7 & 80 & 179\\
Breast cancer (2 classes) &  \ 0 &  \ 0 &\ 9 & 0 & \ 0  & \ 43 \\
Breast cancer (3 classes) &  \ 0 &  \ 0 &\ 3 & 0 & 51 & 246\\
Adenocarcinoma &  \ 0 &  \ 0 &  \ 3 & 0 & 45 & 270\\
NCI 60 &  \ 0 &  \ 0 &  \ 0 & 0 & \ 0 & \ \ 6\\
\hline
\end{tabular}

\caption{\label{table-boot}Stability of results using the bootstrap, with 200
  bootstrap iterations. Values shown are the number of genes that are returned,
  as members of a signature component, in at least those many bootstrap runs.}

\end{center}
\end{table}

\vspace{70pt}
\newpage


\end{document}